\documentclass{article}
\usepackage{arxiv}

\usepackage[utf8]{inputenc} 
\usepackage[T1]{fontenc}    
\usepackage{hyperref}       
\usepackage{url}            
\usepackage{booktabs}       
\usepackage{nicefrac}       
\usepackage{microtype}      
\usepackage{lipsum}		
\usepackage{graphicx}
\usepackage{natbib}
\usepackage{doi}
\usepackage{mwe} 
\usepackage{algorithm}
\usepackage{algorithmic}
\usepackage{fancyref}
\usepackage{rotating}
\usepackage{ bbold }
\usepackage{pifont}
\usepackage{array}
\usepackage{comment}
\usepackage{xcolor}
\usepackage{amsmath,amsfonts,amssymb}
\renewcommand{\shorttitle}{Cackowski \textit{et~al.}}

\newcolumntype{P}[1]{>{\centering\arraybackslash}p{#1}}

\definecolor{newcolor}{rgb}{0.8,0.349,0.1}
\definecolor{mygreen}{RGB}{0, 255,0}

\title{ImUnity: a generalizable VAE-GAN solution for multicenter MR image harmonization}

\author{ \href{https://orcid.org/0000-0001-6647-8419}{\includegraphics[scale=0.06]{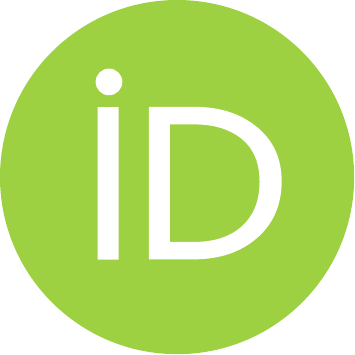}\hspace{1mm}Stenzel Cackowski}\thanks{Universit\'e Grenoble Alpes, Inserm U1216, Grenoble Institut for Neurosciences, 38700 La Tronche, France}\\
	\texttt{stenzel.cackowski@univ-grenoble-alpes.fr} \\
	\And
	\href{https://orcid.org/0000-0002-4952-1240}{\includegraphics[scale=0.06]{images/orcid.pdf}\hspace{1mm}Emmanuel L. Barbier}\footnotemark[1]   \thanks{Corresponding author} \\
	\texttt{emmanuel.barbier@univ-grenoble-alpes.fr} \\
    \And
	\href{https://orcid.org/0000-0003-2747-6845}{\includegraphics[scale=0.06]{images/orcid.pdf}\hspace{1mm}Michel Dojat}\footnotemark[1] \\
	\texttt{michel.dojat@univ-grenoble-alpes.fr} \\
	\And
	Thomas Christen\footnotemark[1] \\
	\texttt{thomas.christen@univ-grenoble-alpes.fr} \\
}

\begin{document}
\maketitle

\begin{abstract}
ImUnity is an original deep-learning model designed for efficient and flexible MR image harmonization. A VAE-GAN network, coupled with a confusion module and an optional biological preservation module, uses multiple 2D-slices taken from different anatomical locations in each subject of the training database, as well as image contrast transformations for its self-supervised training. It eventually generates ‘corrected’ MR images that can be used for various multi-center population studies. Using 3 open source databases (ABIDE, OASIS and SRPBS), which contain MR images from multiple acquisition scanner types or vendors and a large range of subjects ages, we show that ImUnity: (1) outperforms state-of-the-art methods in terms of quality of images generated using traveling subjects; (2) removes sites or scanner biases while improving patients classification; (3) harmonizes data coming from new sites or scanners without the need for an additional fine-tuning and (4) allows the selection of multiple MR reconstructed images according to the desired applications. Tested here on T1-weighted images, ImUnity could be used to harmonize other types of medical images.

\end{abstract}

\keywords{Brain\and Deep Learning \and Adversarial Network \and Machine Learning \and Self-supervised learning \and Radiomic features}

\section{Introduction}
\label{sec:intro}
Magnetic Resonance (MR) data acquired from the same patient but at different acquisition sites often lead to different MR images. This is due to the qualitative nature of the acquisitions which produces weighted images (such as T1w or T2w) that are sensitive to technical choices (hardware, sequence parameters) as well as scanner artifacts. Consequently, pooling images from multi-center MR studies in order to approach a particular clinical or biological question does not guarantee an increase in statistical power because of a parallel increase in non-biological variance. These unwanted variations in image intensities also prevent large dissemination of machine learning tools that are trained on a specific site and may not generalize their model to other image providers \citep{liu_ms-net_2020}. 

Several solutions have been proposed in the last decade to harmonize data coming from multi-site or multi-scanner MR studies. Their goal is to remove confounding site, scanner or protocol effects, while preserving the biological information contained in the images. Classical post-processing steps such as standardization, global scaling \citep{fortin_harmonization_2017} or intensity histogram matching \citep{shinohara_statistical_2014} have been shown to reduce the influence of site or scanner biases. However, they also tend to remove informative local intensity variations. Statistical techniques, where image intensity and datasets bias are modeled in every voxel, have been proved to be more efficient. Ravel \citep{fortin_removing_2016}, ComBat \citep{fortin_harmonization_2017}, refined ComBat versions (\citet{pomponio_harmonization_2019}, \citet{beer_longitudinal_2020}) or dictionary learning \citep{stjean_harmonization_2020} methods have been successfully used to analyze harmonization impacts on diffusion MRIs or longitudinal structural sequences. However, it can be noted that these techniques need to be adjusted every time new sites or scanners provide images to the database. Moreover, the same clinical individual information (such as patient age, sex, etc.) needs to be available in every center of the database. More recently, deep-learning models such as CycleGAN \citep{zhu_unpaired_2018}, Deep-Harmony \citep{dewey_deepharmony:_2019} or Calamity \citep{zuo_information-based_2021} have shown encouraging results for structural MR image (T1w, T2w or FLAIR) harmonization but have also shown limitations. Briefly, CycleGAN, which consists in two Generative Adversarial Networks (GANs) working together, is restricted to the harmonization of 2 sites, and needs to be fine-tuned for every pair of sites. Deep-Harmony, a U-Net \citep{ronneberger_u-net_2015} harmonization adaptation network, has the disadvantage of requiring traveling subjects (subjects who have been scanned successively at different sites or scanners) for all scanners/sites for its training, a condition barely met in practice even in prospective studies. Similarly to CycleGAN, it is limited to two sites and needs to be fine-tuned. Calamity, an unsupervised deep-learning method, needs two different MR sequences as inputs for every subject and needs fine-tuning when data from new sites are considered. Finally, \citet{dinsdale_deep_2020} and \citet{guan_multi-site_2021} have proposed to include unlearning modules or domain discriminators directly into their classification networks. As such, they learn how to remove datasets biases during their analyses without reconstructing harmonized MR images. They have shown improvements in brain tissues segmentation and brain disorder classification after harmonization. However, these techniques clearly require to be trained for every new clinical question while the former approaches harmonize data once for all.

We propose in this paper a new type of harmonization method, called ImUnity, based on deep-learning, which extends previous techniques to offer a fast and flexible harmonization solution. ImUnity generates ‘corrected’ MR images that can then be utilized for various population imaging studies. To avoid the need for traveling subjects or multiple MR sequences in the database, our self-supervised Variational AutoEncoder (VAE-GAN) architecture uses for its training multiple slices from the same individual and randomized image contrast transformations. It also unlearns center bias using a confusion module connected to its bottleneck while an optional biological module can ensure that clinical features are preserved in the latent space. Once trained, this architecture should allow data coming from new sites or scanners to be harmonized without the need for fine-tuning. The architecture also allows estimates towards multiple target sites and then, users can choose multiple MR image reconstructions according to the chosen target domain (site or scanner).

To evaluate the efficiency and flexibility of our harmonization tool, we tested the approach using 3 open source databases that contain images from multiple acquisition sites, scanner vendors or strength of magnetic fields, and a large range of patients ages. For most of the experiments, ImUnity was trained using data from only one of the databases and then applied to the other two to evaluate generalisation of the model. Quality of the reconstructed images, capacity of removing site or scanner bias and ability to classify patients were evaluated after data harmonization.  
  
\begin{table*}[!t]
\centering
 \begin{tabular}{|p{30mm}|p{30mm}|p{25mm} |p{25mm} |p{25mm}|} 
 \hline
  & Travelling subjects & Fine-tuning for new clinical question & Fine-tuning for unseen sites & Max. number of target sites \\
\hline
 \citet{zhu_unpaired_2018} (CycleGAN) & \textcolor{mygreen}{not required} & \textcolor{mygreen}{not required} & required & N = 2\\
 \hline
 \citet{dewey_deepharmony:_2019} (Deep-Harmony) & required & \textcolor{mygreen}{not required} & required & N = 2 \\
 \hline
 \citet{zuo_information-based_2021} (Calamity) & \textcolor{mygreen}{not required} & \textcolor{mygreen}{not required} & required & N = number of training sites\\
 \hline
 \citet{dinsdale_deep_2020}, \citet{guan_multi-site_2021} & \textcolor{mygreen}{not required} & required & \textcolor{mygreen}{not required} & \textcolor{mygreen}{N $>$ number of training sites} \\
 \hline
 This study (ImUnity) & \textcolor{mygreen}{not required} & \textcolor{mygreen}{not required} & \textcolor{mygreen}{not required} & \textcolor{mygreen}{N $>$ number of training sites} \\
 \hline
\end{tabular}
\caption{Versatility of deep-learning harmonization models}
\label{table:requirements}
\end{table*}

\section{Materials and methods}
\label{sec:m&m}

\subsection{Data}
\label{sec:data}
We used three open-source databases: (1) \href{http://fcon_1000.projects.nitrc.org/indi/abide/}{ABIDE}, a multi-center project led by \cite{di_martino_autism_2014}, which focuses on Autism Spectrum Disorder (ASD). It gathers more than 1,000 autistic patients and controls. For this study, we used T1-weighted scans from 11 different sites and scanners from 3 different constructors (3T scanners at 10 different sites and one 1.5T scanner at one site). Sites presenting data from a large range of ages (from 6 to 47 years, mean age = 12 years) were selected. In total, 621 T1-weighted scans (309 patients and 312 controls) were collected. (2) \href{https://www.oasis-brains.org/}{OASIS} (\cite{lamontagne_oasis-3_2019}) gathers T1-weighted scans from adult subjects who underwent several MR sessions on 4 different scanners from the same site. We used these traveling subjects (n = 769) to validate the ability of our model to perform multi-scanners harmonization. \\
(3) \href{https://bicr-resource.atr.jp/srpbsts/}{SRPBS} (\cite{tanaka_multi-site_2021}) is a multi-site database gathering multi-disorder subjects. We used 9 healthy adult traveling subjects to validate harmonization results between the different acquisition sites of the database (6 sites, 12 scanners from 3 different constructors). 
Note that OASIS and SRPBS images contain healthy adult brain scans while ABIDE mainly include healthy and pathological infant brain scans, leading to large anatomical differences between images in the databases. 

For each subject in each database, the brain was extracted using Robex \citep{iglesias_robust_2011} and N4Bias \citep{tustison_n4itk_2010} was used to correct for intensity inhomogeneities. MR images were first co-registered to the publicly available and age specific 152-MNI templates \citep{sanchez_age-specific_2012}. Then, White-Stripe normalization \citep{shinohara_statistical_2014} was run to align white matter (WM) peaks between all subjects (each WM peak was aligned to 0.7 after rescaling the whole image between [0:1]). 

After visual inspection to detect images with ROBEX defects or other artifacts, we eventually included 545, 1072 and 81 T1-weighted scans from ABIDE, OASIS and SRPBS databases respectively.

\begin{figure*}[!h]
\centering
\includegraphics[width=\linewidth]{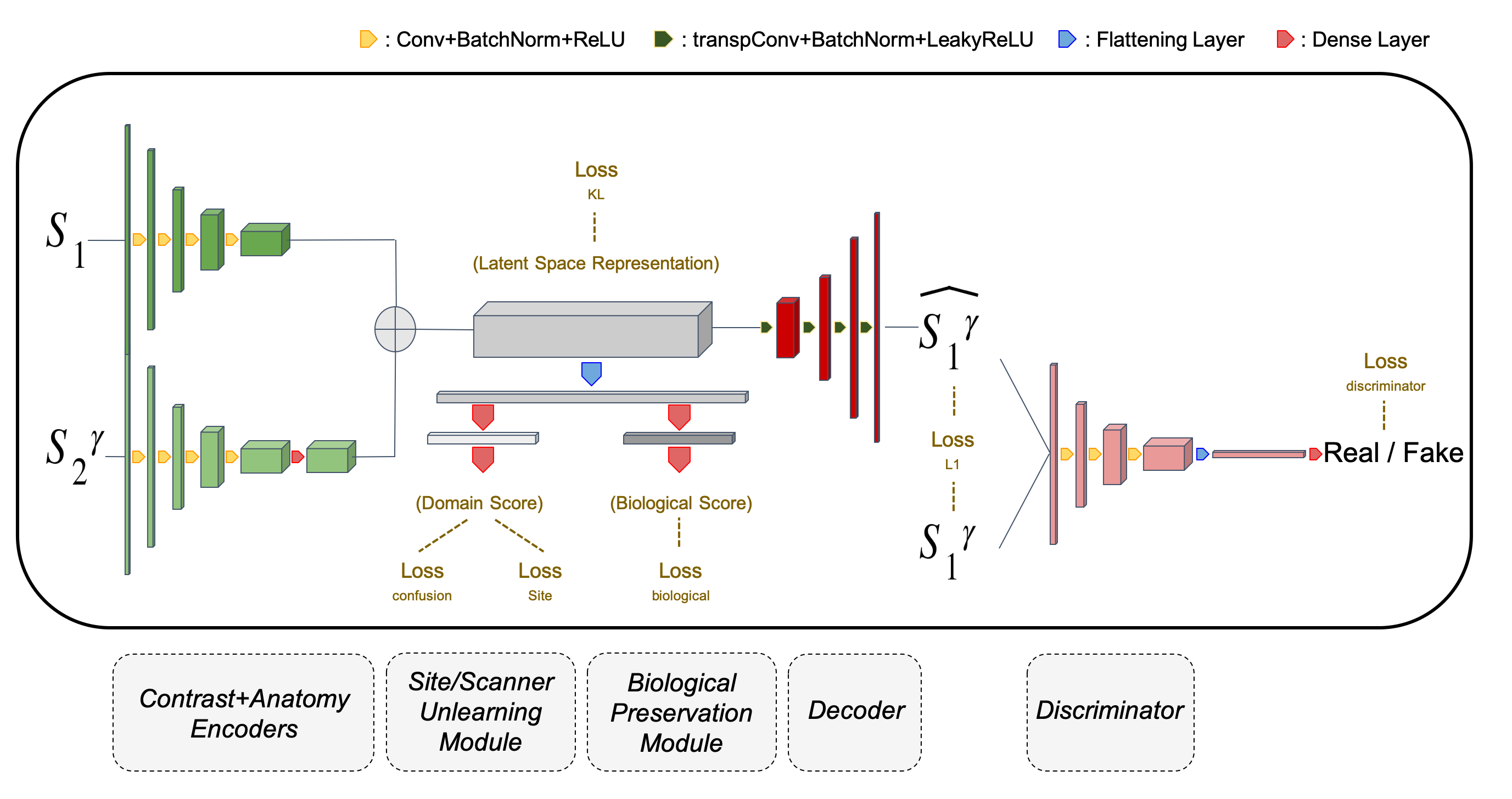}

\caption{ImUnity's architecture. The model involves: a modified VAE generator (\ref{sec:generator}), a CNN discriminator, an additional Site Unlearning module (\ref{sec:unlearning_module}) and an optional Biological module (\ref{sec:bio_module}).}
\label{fig:model}
\end{figure*}

\subsection{ImUnity's model}
The architecture of our model derives from convolutional VAE-GANs and is described in Figure \ref{fig:model}. We adopted adversarial settings to ensure realistic outputs using a classical CNN as discriminator. The generator (here a VAE) learns how to represent input data into a lower dimension latent space (bottleneck). Information is then decoded to generate an output image. Inspired by \citet{dinsdale_deep_2020}, an unlearning center-bias module is connected to the bottleneck to limit the impact of site or scanner information. A biological preservation module can be inserted to maintain biological information in the latent space representation. Technical details are provided below. 

\subsubsection{Modified VAE generator}
\label{sec:generator}
\begin{figure}[!ht]
\centering
\includegraphics[width=0.7\linewidth]{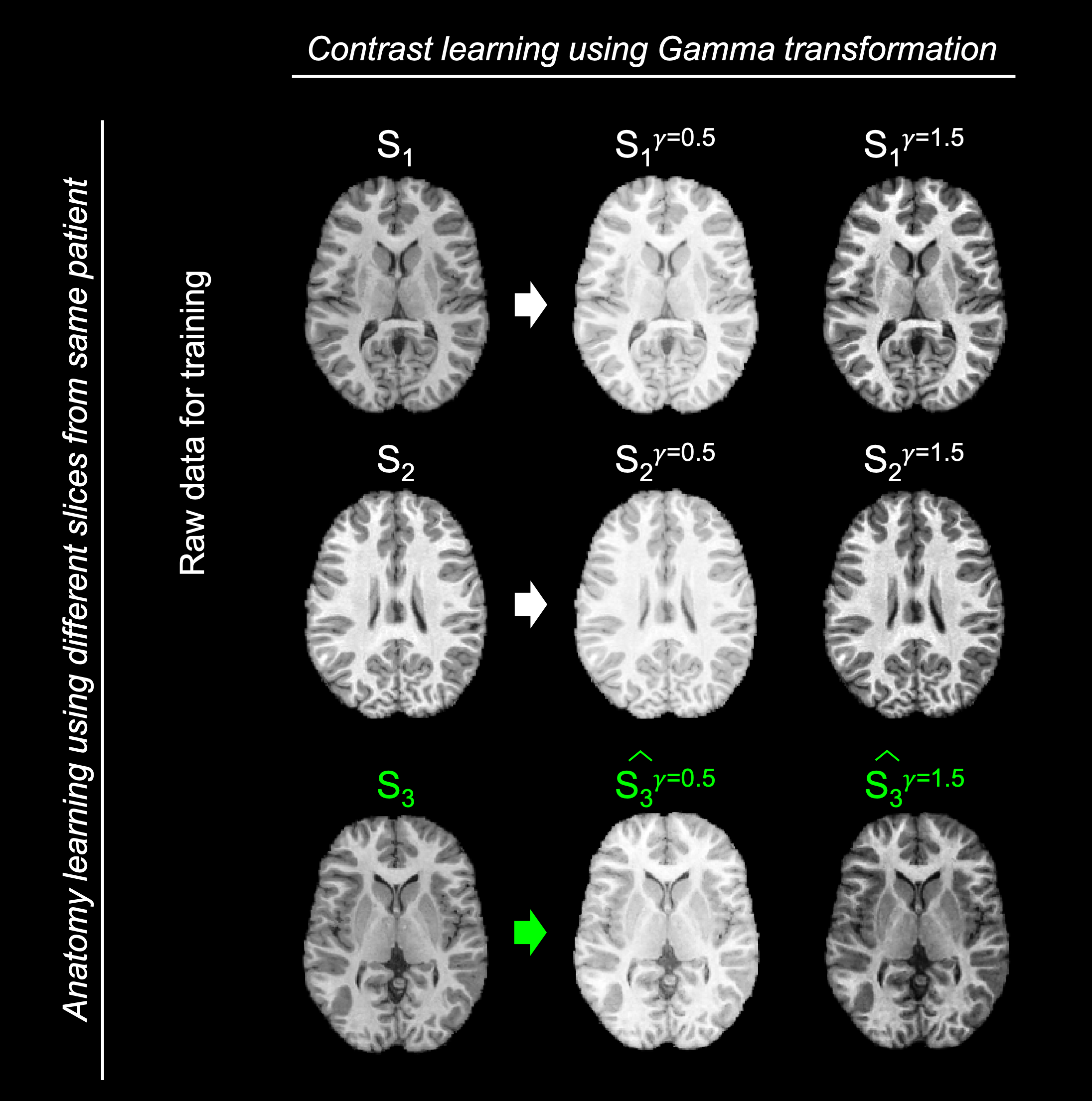}

\caption{Inputs and outputs of the model: Different slices from the same patient are used to encode the anatomical information. Gamma transformations are used to encode the contrast information. Rows represent different anatomical slices ($S_1$, $S_2$ and $S_3$) taken from the same subject. The first two rows present Gamma transformations used to train the model ($S_1^{\gamma}$ and $S_2^{\gamma}$). The last raw shows model outputs ($\hat{S_3^{\gamma}}$) for estimated Gamma transformations for the slice $S_3$ not present in the training set. From left to right columns: original slices, Gamma modified slices with parameter 0.5, Gamma modified slices with parameter 1.5.}
\label{fig:gamma_slices}
\end{figure}

Inspired by \citet{zuo_information-based_2021}, our generator takes two 2D-structural images as input, randomly taken at two different locations in the 3D-MR stack of images of each subject to consider. The first ($S_1$) image is used by the first CNN to encode the 'anatomical' information using only convolutional filters to ensure preservation of spatial information. The second image ($S_2$), different from $S_1$ because randomly taken in another part of the brain, provides the initial ‘contrast’ information. Thus, $S_1$ and $S_2$ have different anatomy (different location in the brain) but similar contrast (same scan). $S_2$ contrast is modified using a gamma function with a random ‘gamma’ parameter sampled uniformly between 0.5 and 1.5 for each new input image. This modified $S_2^{\gamma}$ slice is used as input to a second CNN to encode the ‘contrast information’ followed by a dense layer to reduce spatial information. An example of different gamma transformations applied to MR brain scans from the same subject is given in Figure \ref{fig:gamma_slices}. Once encoded, the two independent representations of $S_1$ and $S_2^{\gamma}$ are concatenated to give a latent space representation which is decoded to create the output $\hat{S_1^{\gamma}}$ using transposed convolutional filters.
 Eventually, this output is compared to the reference gamma modified slice $S_1^{\gamma}$. Note that this generator is trained in a self-supervised fashion as it generates its own outputs. It does not require additional information such as scanner, center or biological information.

\subsubsection{Site/Scanner-bias unlearning module}
\label{sec:unlearning_module}
To ensure the task of "removing site or scanner bias", a module is directly connected to the encoders' outputs (latent space representation of inputs). The module can be seen as a domain (site or scanner) discriminator and is trained independently from the encoder to predict the scan's origin based on the latent space representation. On the other side, the encoder is trained in an adversarial fashion. A confusion loss is used to unlearn domain information. This principle has been introduced in the field of domain adaptation by \citet{ganin_domain-adversarial_2016} and has been adapted to medical imaging studies by \citet{dinsdale_deep_2020}. Originally, the module was incorporated directly in the model to unlearn datasets bias and to improve predictions. Here, it is used in the bottleneck as a "datasets bias filter", forcing the encoder to learn a domain-invariant data representation. Thus, the generator learns a shared latent space that encodes all information needed to generate harmonized scans. Note that no skip connection is used in the generator architecture, preventing the presence of datasets bias in the final output. The loss function for the site/scanner unlearning module is: 
\begin{equation}
\label{eq:l_site}
l_{site}(P,Y) = -(1/N)\sum_{i=1}^N\sum_{s=1}^S \mathbb{1}(y_i=s)log(p_i^s)
\end{equation}
While the confusion loss used in the encoders' training is :
\begin{equation}
\label{eq:l_conf}
l_{confusion}(P) = -(1/N)\sum_{i=1}^N\sum_{s=1}^S log(p_i^s)/S
\end{equation}
Here $P = [p_1; ...; p_S]$ is the softmax output from the module, corresponding to the probability to belong to different sites (1, ..., S) , Y is the ground truth site affiliation vector, and N is the sample size. 

\subsubsection{Biological preservation module}
\label{sec:bio_module}
An optional module ensures the "preservation" of biological information. It acts as a classifier of available biological information. For instance, features such as age or presence of diseases can be introduced. Contrary to the unlearning module, the encoder is trained to minimize its loss function. This module is not mandatory, and a fully self-supervised learning model can be adopted if it is turned off. The loss function of the biological preservation module for our particular application using the ABIDE database is:
\begin{equation}
\label{eq:l_bio}
\resizebox{0.91\hsize}{!}{
$l_{biological}(P, Y) = -(1/N)\sum_{i=1}^{N}\sum_{f \in features}\sum_{i=1}^{N} y_i^f log(p_i^f) + (1 - y_i^f) log(1 - p_i^f)$
}
 \end{equation}

Here P represents module's predictions for biological features of interest, N is the sample size while Y is the ground truth vector. Note that in this study the binary cross entropy formulation was used for the loss function because only two features (age and patient status, i.e. ASD) were considered.

\subsection{Training}
Training the model involves several independent steps, due to the adversarial context and the use of the additional modules. \\
\begin{itemize}
\item Training the discriminator consists in minimizing binary cross-entropy $l_{discriminator}$ between its predictions and the labels corresponding to the nature of the inputs (real or fake). Adversarially, the generator learns how to maximize this loss function, forcing the generation of realistic outputs.\\

\item Training the site/scanner unlearning module consists in minimizing the categorical cross-entropy (eq. \ref{eq:l_site}) between its predictions and the site-affiliation labels. Adversarially, the generator is trained to minimize the confusion loss (eq. \ref{eq:l_conf}). It forces a site and scanner invariant representation of the dataset in the latent space, leading to uniform outputs of the unlearning module. \\

\item In our ABIDE experiment, training the biological preservation module consisted in minimizing binary cross-entropy losses associated with each biological feature taken into account (here sex and patient status). Unlike the previous module, the loss $l_{biological}$ was directly integrated in the generator. This ensures the conservation of biological features in the latent space.\\

\item Besides previous loss functions involved in training the generator, a $l_1$ loss function is used to ensure a good mapping between input $(S_1;S_2^{\gamma})$ and the generated output $\hat{S_1^{\gamma}}$ ($l_1 = mean(|\hat{S_1^{\gamma}} - S_1^{\gamma}|)$). Moreover, the use of the Kullback-Leibler divergence $l_{KL}$ between features distributions and a Gaussian distribution ensures a dense data representation in latent space. The global generator's loss function to minimize is therefore:\\

\begin{equation}
\label{eq:global_loss}
l_{generator} = - \lambda_1l_{discriminator} + \lambda_2l_{confusion} + \lambda_3l_{biological} + \lambda_4l_1 + \lambda_5l_{KL}
\end{equation}

Where $\lambda$ factors control the relative contribution of each loss. In our study we used : $\lambda_1 = 1$; $\lambda_2 = 1$; $\lambda_3 = 1$; $\lambda_4 = 100$; $\lambda_5 = 10^{-3}$ found empirically.\\
\end{itemize}

\subsection{Experiments}
 The datasets extracted from the three databases were used to evaluate different aspects of our model. Impact on image quality in multi-site or multi-scanner harmonization was assessed using data from traveling subjects (ground truth) from the OASIS and SRPBS datasets. Ability to remove site information was evaluated using the ABIDE dataset. Finally, benefits of harmonization between data provider centers were assessed using the predictions of autism disorder in children from the ABIDE dataset. 
To demonstrate the flexibility of ImUnity, all experiments were performed with the same model trained on data coming from the ABIDE database, unless specified. OASIS and SRPBS were then used for the validation parts only. The model was trained on 2D axial slices with at least 1\% of brain tissue voxels. Training was run on a Nvidia GeForce 2080 RTX for 2000 epochs using a learning rate of $10^{-4}$ and Adam optimizer.\\ 

\subsubsection{Experiment 1 : Harmonization on traveling subjects (OASIS+SRPBS)}
\label{sec:exp1}
We first evaluated the ability of our model to transform images from one domain (site or scanner) to their equivalent in another domain. As SRPBS and OASIS databases contain traveling subjects, ground truth was available to assess ImUnity performances. In practice, one domain (acquisition site or scanner) was first selected as the reference for every subject. Individual scans were co-registered to their equivalent in the reference domain (to avoid variations between acquisitions due to movement). Then, all the images were transformed by the model into the reference domain. During this step, slices to be harmonized (anatomy) were fed to the model alongside the corresponding computed contrast slice from the reference domain. Finally, results obtained after transformation were compared to the ground truth, i.e. images acquired in the reference domain (traveling subject). Visual verification, histograms of image intensities, and the Structural Similarity Index Metric (SSIM, \citet{wang_multiscale_2003}) were used to assess image likeness. The same model, trained on ABIDE data, was used for every site/scanner of the two other databases to evaluate the ability of ImUnity to generalize to sites never seen before. This experiment also evaluates ImUnity's versatility, either for the source domain or for the target domain (last two columns of Table \ref{table:requirements}).\\

\subsubsection{Experiments 2 : Harmonization's effects on sites classification (ABIDE)}
\label{sec:exp2}
The second experiment evaluated the ability to detect the origin of data before and after harmonization. As no ground truth was available for this experiment, we considered harmonization impacts on classification algorithms. Standard Support Vector Machine (SVM) with a radial basis function kernel was used to classify ABIDE data. The classifier worked on all radiomic features (N=101) extracted using the pyradiomics python API \citep{van_griethuysen_computational_2017}. These features aim to represent different aspects of MRI images such as shape, contrast or texture and are known to be sensitive to site effects \citep{orlhac_validation_2019}. The most 'correlated features’ with sites affiliations before harmonization were selected for classification using Pearson tests (ran independently for each feature) using $10^{-3}$ as p-value threshold (30 features in total). Accuracy and Area Under Curve (AUC) of the Receiver Operating Characteristic (ROC) curve were used to evaluate the specificity and sensitivity of the site classifier. \\

\subsubsection{Experiments 3 : Harmonization's effects on autism syndrome disorder prediction (ABIDE)}
\label{sec:exp3}
Similarly to Experiment 2, Experiment 3 evaluated the ability of our classifier to detect patients with ASD from the ABIDE database, before and after harmonization. Here, results were obtained following a 10 fold-cross-validation procedure. The same trained model was used for different number of sites (and for different combinations of sites) included in the ABIDE database. \\

\begin{figure}[!ht]
\centering
\includegraphics[width=0.7\linewidth]{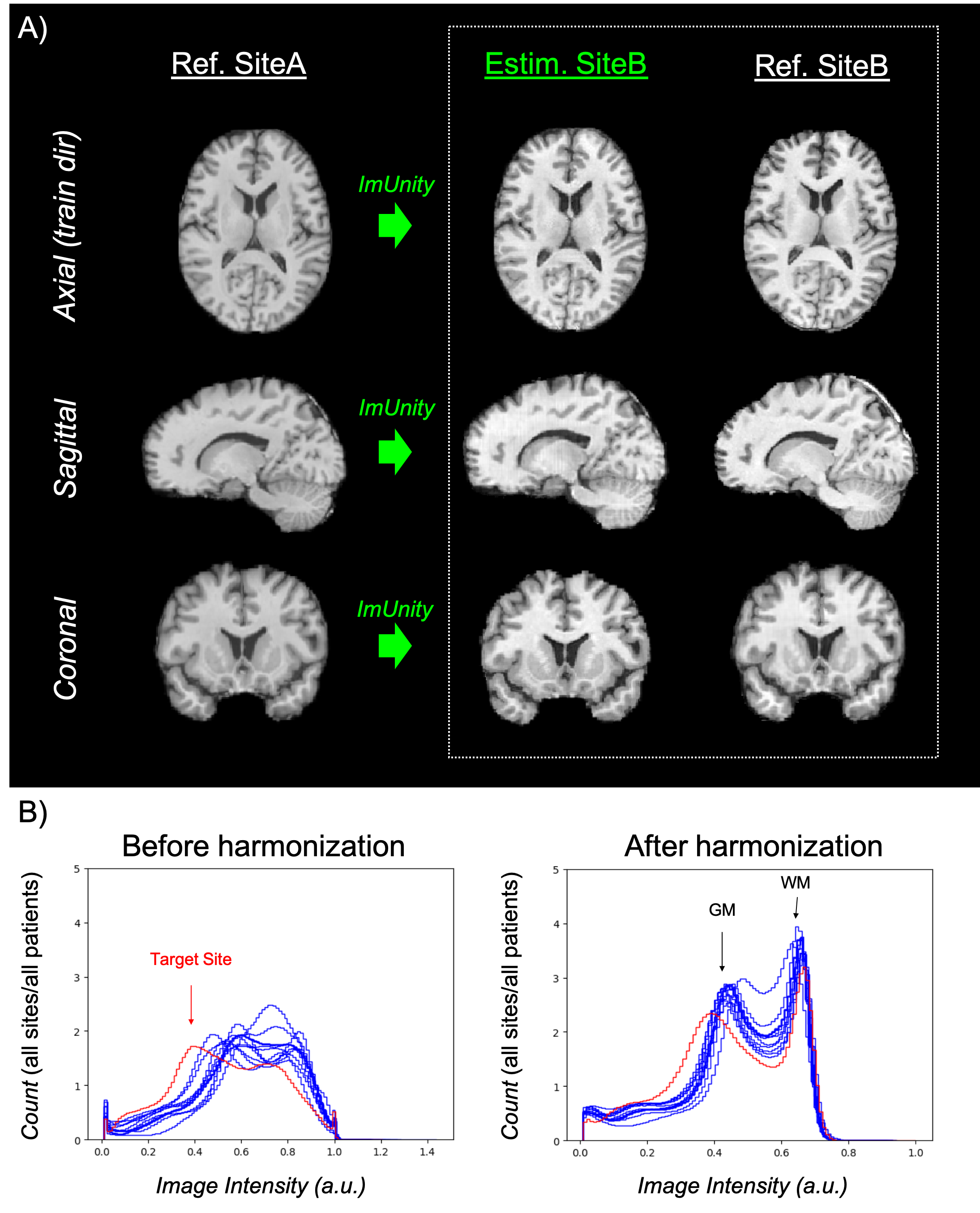}

\caption{Harmonization result on travelling subjects from the SRPBS database. A) Left: 3D images from one patient (axial, sagittal, coronal views) acquired in site A before harmonization, Middle: ImUnity's harmonization to fit with acquisition at site B, Right: image acquired at site B (ground truth). Note that the ImUnity model was trained on axial slices only. B) Images intensity distributions (all patients) before (Left) and after ImUnity's harmonization (Right). The red histogram corresponds to the site taken as reference for the harmonization process (target site).\\
GM = gray matter; WM = white matter}
\label{fig:SRPBS_results}
\end{figure}

\section{Results}
\label{sec:res}

\textit{Experiment 1}: Figure \ref{fig:SRPBS_results}-A shows the results obtained for one traveling subject from the SRPBS database (\ref{sec:exp1}). Images are shown for one acquisition at site (A) before harmonization (\ref{fig:SRPBS_results}-A, left), corrected by ImUnity to fit with acquisition at site B (\ref{fig:SRPBS_results}-A middle) and the corresponding ground truth acquired at site B (\ref{fig:SRPBS_results}-A right). One can notice the difference in image contrast between the 2 sites, highlighting the need for image harmonization, as well as the visual similarity between the harmonized image and the ground truth. It is interesting to observe that the anatomical structures of the input contrast reference are not propagated through the model, which explains small anatomical differences (e.g. superior sagittal sinus) between the model estimates and the ground truth. It is also worth noting that although the model was trained on 2D-axial slices, the 3D reconstructions of the estimates are of high quality in each orientation. Figure \ref{fig:SRPBS_results}-B shows the effects of ImUnity's harmonization on image intensity distributions for all selected subjects from the SRPBS database. The model was used to harmonize every image to a target site (indicated in red). An alignment of histograms can be clearly observed after harmonization, with both gray and white matter peaks shifted. Changes in intensity distribution of the site of reference are due to pre-processing (see details in Supplementary Material (SM), Figure \ref{appendix:fig:wm_impacts}, top row). Images obtained after ImUnity's harmonization of the OASIS datasets are provided in SM, Figure \ref{appendix:fig:OASIS_results}.\\

\begin{table}[!ht]
\centering
  \begin{tabular}{|c|p{32mm} | p{35mm} |c|} \hline
Task & \multicolumn{2}{c|}{Multi-scanner harmonization.$\newline$} & Multi-site$\newline$harmonization \\
\hline
 Dataset & OASIS $\newline$scanner F $\rightarrow$ scanner E & OASIS $\newline$all scanners $\rightarrow$ scanner E & SRPBS $\newline$all sites $\rightarrow$ UTO site \\
\hline\hline
 Raw data & 0.871 $\pm$ 0.045 & 0.845 $\pm$ 0.059 & 0.853 $\pm$ 0.021 \\
 \hline
 \citet{zhu_unpaired_2018} CycleGAN$^*$ & 0.873 $\pm$ 0.046 & - & - \\
 \hline
 \citet{zuo_information-based_2021} Calamity$^*$ & 0.884 $\pm$ 0.046 & - & - \\
 \hline
 ImUnity$^{\#}$ & {0.920 $\pm$ 0.024} & {0.919 $\pm$ 0.023} & \textbf{0.907 $\pm$ 0.024} \\
 \hline
 ImUnity$^*$ & \textbf{0.943 $\pm$ 0.003} & \textbf{0.943 $\pm$ 0.003} & {0.893 $\pm$ 0.025}\\
 \hline
  ImUnity$^+$ & 0.824 $\pm$ 0.064 & 0.827 $\pm$ 0.061 & 0.865 $\pm$ 0.019\\
 \hline
\end{tabular}
\caption{SSIM in travelling subjects for multi-scanner (OASIS database) and multi-site (SRPBS database) harmonization. Results are compared to the literature when available.\\
 $*$ : Model trained on OASIS database (n=1072); $\#$ : Model trained on ABIDE database (n=545); $+$ : Model trained on SRPBS database (n=81)}
\label{table:ssim_metrics}
\end{table}

Quantitative results obtained with the SSIM metric in all traveling subjects are summarized in Table \ref{table:ssim_metrics}. Both multi-site (SRPBS) and multi-scanner (OASIS) experiments are shown. For the latter, results from the literature are also given for reference. It can be seen that ImUnity increases the structural similarity in all cases and provides better performances compared to other deep learning approaches. Moreover, results from multi-scanner harmonization show that ImUnity performs well independently of the chosen reference domain. The last 2 lines present results obtained after training ImUnity on OASIS (to better match literature protocols) and SRPBS data. These models were used to harmonize OASIS as well as SRPBS data. \\

\begin{figure}[!ht]
\centering
 \includegraphics[width=0.5\linewidth]{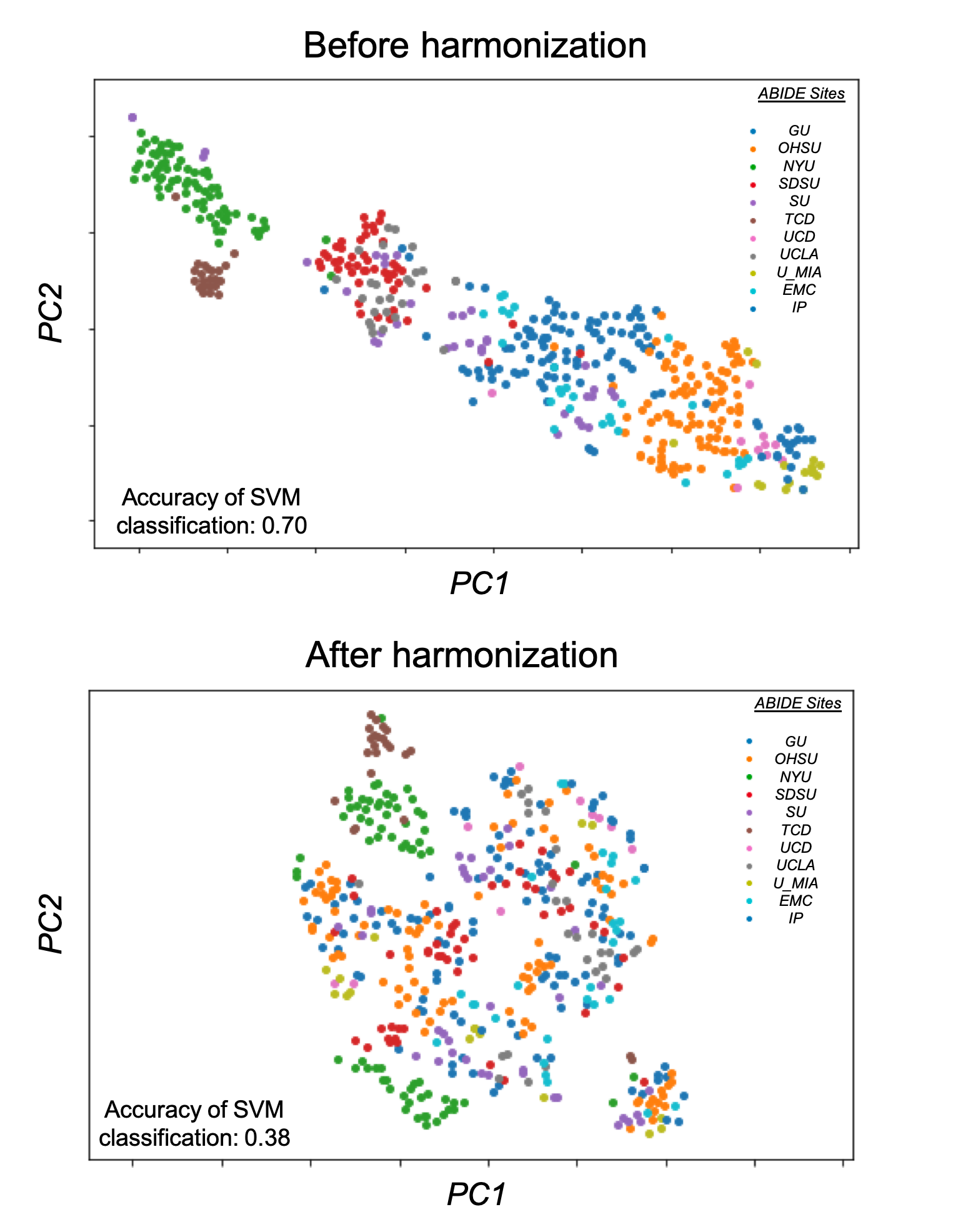}
\caption{Harmonization effects on ABIDE sites classification. 2D scans representation of ABIDE database using tSNE reduction algorithm are presented before and after harmonization. Colors correspond to different sites. The separation of data provided by the different sites is clearly more difficult after harmonization.}
\label{fig:sites_classification}
\end{figure}

\textit{Experiment 2}: Figure \ref{fig:sites_classification} shows ImUnity's harmonization effects on site classification on the ABIDE datasets (\ref{sec:exp2}) using tSNE \citep{maaten_visualizing_2008}, a dimension reduction algorithm, on radiomic features. Before harmonization, the presence of site clusters is clear. Once the data is harmonized using ImUnity, the points are shuffled and the accuracy of the SVM site prediction decreases from 0.70 to 0.38 (before and after harmonization respectively). This confirms the removal of site bias by ImUnity as the classifier is no longer able to correctly separate the sites. Additional results on the influence of the preprocessing step on sites classification are provided in SM Fig. \ref{appendix:fig:wm_impacts}.\\

\begin{figure}[!ht]
\centering
 \includegraphics[width=0.5\linewidth]{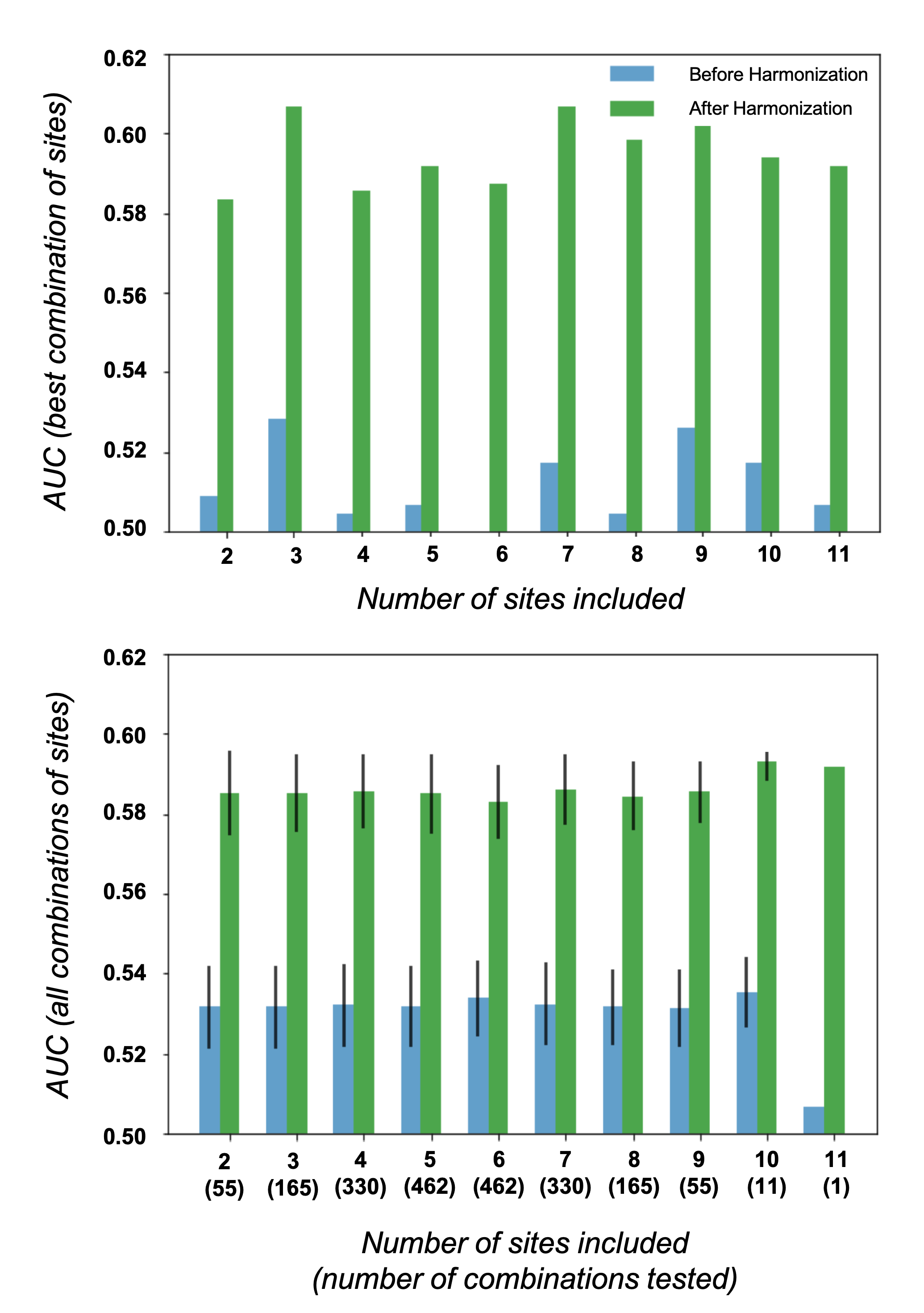}
\caption{Harmonization effects on ABIDE patients classification. AUC metric for classification of patients with autism spectrum disorder (ASD) using SVM and extracted radiomic features (Experience 3, \ref{sec:exp3}), are shown for different numbers of sites included in the database (from 2 to 11 sites). \\Top row: Results obtained from the best (largest change in AUC before and after harmonization) combination of sites. Bottom row : Average and standard deviation of AUC estimates for all combination of sites. Same trained model and harmonized data were used for different site combinations.}
\label{fig:clinical_impact}
\end{figure}

\textit{Experiment 3}: Figure \ref{fig:clinical_impact} shows the capacity of our model to improve ASD prediction from the ABIDE datasets. Here, we used the same trained model to test the influence of different numbers of sites included in the database (from 2 to 11) as well as different combinations of those sites (for example 55 combinations of 2 sites taken among the 11 sites available). In every case, we observed a clear improvement of classification of autistic patients after harmonization as shown by increases in AUC provided by the SVM classifier. We show the results obtained with the best combination of sites as well as average and standard deviation of AUC with all combinations of sites. The preprocessing also has a positive impact on the prediction as shown in SM (Fig. \ref{appendix:fig:wm_impacts}, bottom row).\\

\section{Discussion}
\label{sec:discussion}
We have presented ImUnity, an original harmonization tool for multi-center MRI databases. ImUnity shows high performances in term of quality of the generated harmonized images, as well as clear removal of the idiosyncratic bias attached to site-dependent image acquisition conditions. Moreover, the performed experiments clearly demonstrate ImUnity's versatility. By training ImUnity's model on datasets extracted from one database (here ABIDE) and looking at images harmonized from traveling subjects provided by two different databases (here OASIS and SRPBS) , we showed that ImUnity did not require new training phase to generalize to unseen sites or scanners (see Fig. 3). The performances were maintained independently of the site selected as reference (see Table \ref{table:ssim_metrics}). While the model was trained on ABIDE data only, it provided better results than the state-of-the-art methods in terms of image quality (+4\%, see Table \ref{table:ssim_metrics}).

The last two rows of Table \ref{table:ssim_metrics} still present SSIM metrics obtained when training ImUnity on the other 2 datasets (OASIS and SRPBS). As no biological features were available in these databases, the biological module was disabled and the model was trained in a self-supervised way. First, we noted additional improvements for scanner harmonization when the model was trained and applied on the same database (here OASIS, +2.5\%). Second, the score obtained for multi-site harmonization (SRPBS) was the highest when trained on ABIDE (N = 545) data (with a slightly better score than with OASIS). It is interesting to observe the impact of the dataset size, the number of site/scanner involved in the training, the use of the biological module and the anatomical differences between datasets (ABIDE mainly contains children data while OASIS and SRPBS focus on adults) on these scores. ABIDE result suggests that anatomical differences could be compensated by a large training dataset presenting more site/scanner variability than OASIS (11 sites for ABIDE vs. 4 sites for OASIS). On the other hand, results from multi-scanners harmonization depict the difficulty of the model trained on SRPBS data to generalize its training to the OASIS data. This indicates an over-fitting effect in this situation, as there was not enough training data (here N=81) and suggests that ImUnity may not be adapted to small sample size scenarios.

In addition to the ability to remove sites or scanners biases, we evaluated the ability to improve subsequent clinical data analysis. The performances of ASD classification obtained on the ABIDE database were improved by ImUnity's harmonization of the datasets even when a simple classification procedure (classical SVM and radiomic features derived from structural MRI) was used. It is interesting to note that several studies have already tackled the issue of ASD patients classification using the ABIDE database \citep{gao_multisite_2020, sherkatghanad_automated_2019}. They used either more complex features (e.g. morphological brain networks) or more complex classifiers (Random Forest, Resnet, adapted deep-learning architecture, etc.) than in our study. However, the reported AUC metrics were close to ours (0.67 by \citet{gao_multisite_2020} and 0.75 by \citet{sherkatghanad_automated_2019}). They also noted the difficulty of obtaining better results when more acquisition sites where included in the study. They did not however mention any harmonization procedure. Thus, we may expect an improvement of ImUnity's performances when introducing more informative radiomic features and choosing a more sophisticated classifier tool.  

Because ImUnity is designed to reconstruct images and to create a new harmonized database, it does not need new training for new clinical or biological questions. Beyond classification, new clinical data investigation should be conducted with ABIDE (or other multi-center clinical databases) to have better understanding on the impact of our method on clinical research studies.

Like the majority of deep networks used for medical image analysis, the MR images used as inputs of our network were first pre-processed for intensity normalization, co-registration or brain extraction. Usually, the impact of these transformations is not examined in harmonization studies. In Fig. \ref{appendix:fig:wm_impacts} (SM), we have highlighted the fact that these steps were already able to remove some of the sites and scanners biases with positive impacts on intensity distributions across sites or patients classification. Intrinsically, the use of White-Stripe normalization \citep{shinohara_statistical_2014} forces the alignment of intensity distributions. Yet, a perfect alignment is not the ultimate goal of harmonization as we also seek to preserve informative biological variations which vary independently across sites. Eventually, we observed that the best results were obtained for all experiments after the whole ImUnity process, with better intensity distributions alignments, removal of persistent datasets noises, and most importantly improvement in patients classification results. On the opposite, other experiments (not reported) also showed that the VAE-GAN network alone performed poorer when the pre-processing steps were omitted, suggesting that these steps are necessary to simplify the training process and improve generalization of the results.

In our study, we only worked with anatomical T1-weighted images. We showed that a single type of sequence, combined with computed image transformations with the Gamma function, are sufficient to learn contrast mapping. This greatly facilitates the use of our model because of few data requirements (the origin of each scan is the only pre-required information) and the possibility of self-supervised training. Yet, we believe that this approach is not only dedicated to T1 contrast harmonization and can easily be generalized to any MRI sequences. Presently, the model needs to be fine-tuned in order to harmonize a new medical imaging type. It could however be interesting to investigate its capacity to learn how to harmonize multiple sequences at once. This could be done by mixing sequence types in our training dataset and ensuring the conservation of this information by adding a new conservation module in the bottleneck. It could also be interesting to add other types of artificial contrast transformation for our training in order to account for other types of sites or sequences biases. 

\section{Conclusion}

\label{sec:conclusion}

We presented ImUnity, an original and effective tool dedicated to MRI harmonization. Our proposed model derives from the VAE-GAN architecture. It ensures realistic outputs and allows removal of idiosyncratic datasets bias and the preservation of biological information. Our results show that the method reaches state-of-the-art results in term of image quality in traveling patients of the OASIS and SRPBS databases and improves autistic patients classification in the ABIDE database. The proposed model is versatile, requiring only one type of MR sequence without the need of matching subjects, can be generalized to sites unseen during the training phase and can be used to harmonize MR images to different reference domains without a new training phase.

\section{Compliance with ethical standards}

This research study was conducted retrospectively using human subject data made available by the following open sources: \href{http://fcon_1000.projects.nitrc.org/indi/abide/}{ABIDE},  \href{https://www.oasis-brains.org/}{OASIS}, \href{https://bicr-resource.atr.jp/srpbsts/}{SRPBS}. Ethical approval was not required as confirmed by the license attached with the data.

\section{Acknowledgments}

\label{sec:acknowledgments}
Stenzel Cackowski is supported by MIAI@Grenoble Alpes (ANR 19-P3IA-003). The authors declare that they have no known competing financial interests or personal relationships that could have appeared to influence the work reported in this paper.

\bibliographystyle{unsrtnat}
\bibliography{Biblio_Stenzel}

\newpage
\onecolumn
\section*{Supplementary Material}
\appendix

\section{Impact of pre-processing steps}

\begin{figure}[!ht]
\centering
\includegraphics[width=\linewidth]{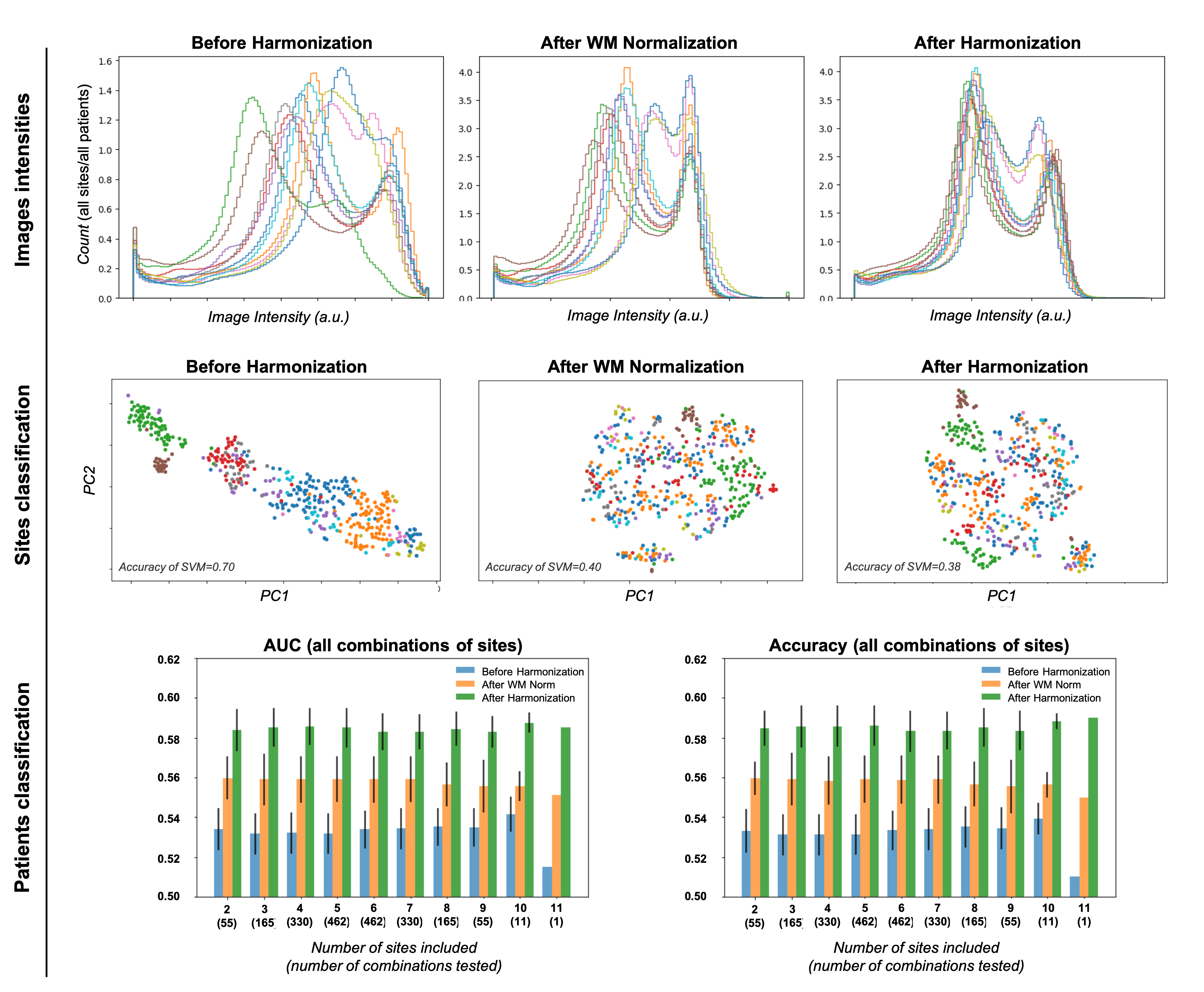}

\caption{Impact of pre-processing steps (N4Biais, White-Stripe normalization) on our different experiments. From top to bottom row: impact on images intensity, impact on sites classification (\ref{sec:exp2}), impact on patients classification (\ref{sec:exp3})}
\label{appendix:fig:wm_impacts}
\end{figure}

\newpage
\section{Scanner harmonization}
\begin{figure}[!ht]
\centering
\includegraphics[width=0.6\linewidth]{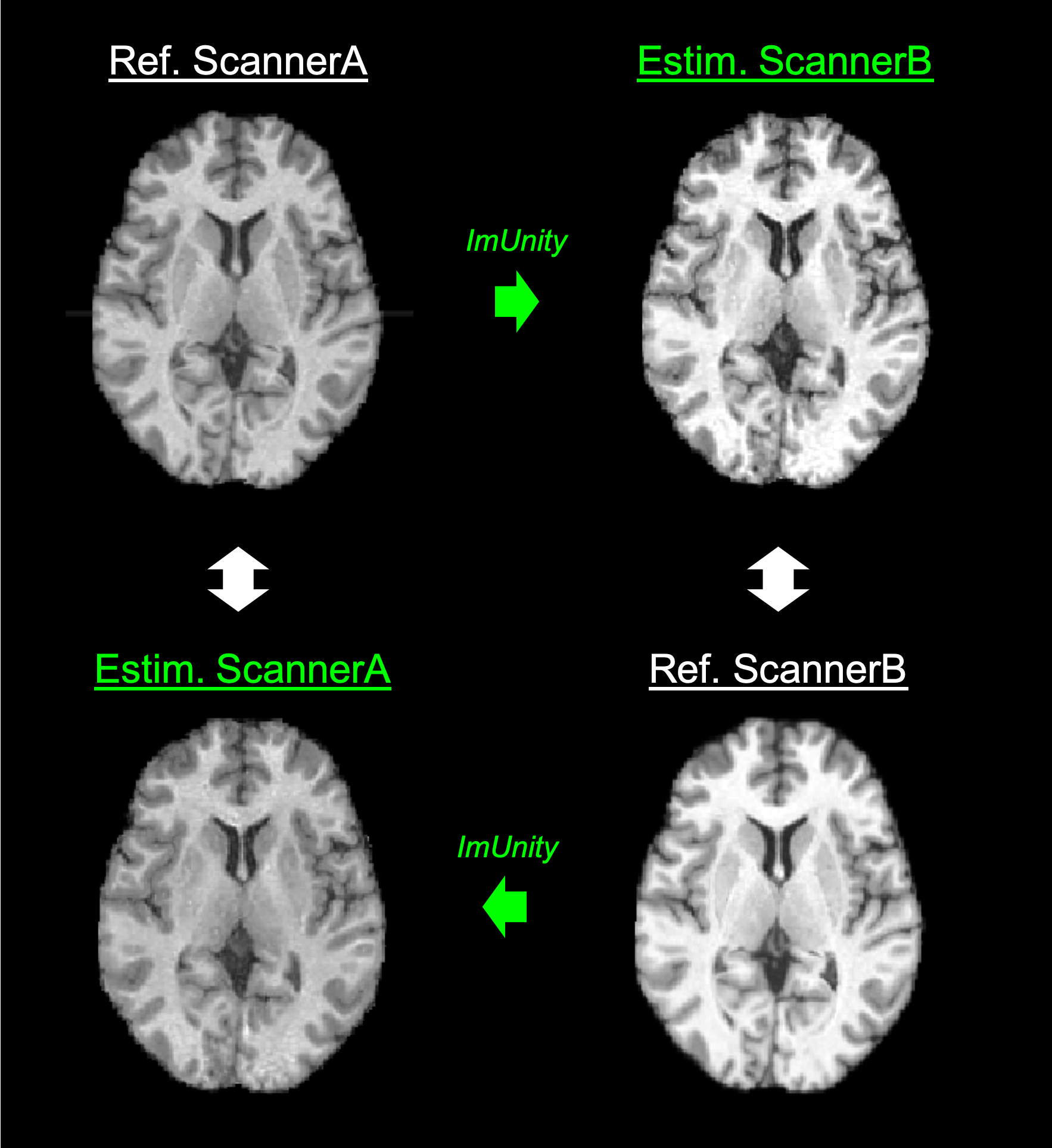}

\caption{Multi-scanner harmonization (\ref{sec:exp1}) results between 2 scanners for the same subject extracted from the OASIS database. ImUnity's model was trained on datasets extracted from the ABIDE database. Top row: Slice acquired at site A (left) and corresponding harmonized image (right) matching acquisitions at site B. Bottom row: Slice acquired at site B (right) and corresponding harmonized image (left) matching acquisitions at site A. Left (resp. right) column allows to visually compare the ground truth and the estimated image for site A (resp. for site B). }
\label{appendix:fig:OASIS_results}
\end{figure}

\end{document}